\documentclass[aps,preprint]{revtex4}

\usepackage{graphics,epsfig,amssymb}

\def\beq{\begin{equation}} 
\def\eeq{\end{equation}} 
\begin{document}

\title{$\alpha$-like quartetting in the excited states of proton-neutron pairing Hamiltonians}

\author{M. Sambataro$^a$ and N. Sandulescu$^b$}
\affiliation{$^a$Istituto Nazionale di Fisica Nucleare - Sezione di Catania,
Via S. Sofia 64, I-95123 Catania, Italy \\
$^b$National Institute of Physics and Nuclear Engineering, P.O. Box MG-6, 
Magurele, Bucharest, Romania}

\begin{abstract}

Previous studies have shown that the ground state of systems of nucleons composed by an equal number of
protons and neutrons interacting via proton-neutron pairing forces can be  described  accurately by a
condensate of $\alpha$-like quartets.  Here we extend these studies to the low-lowing excited
states of these systems  and show that these states can be accurately described by breaking a quartet from the ground state condensate and replacing it with an ``excited"  quartet. This approach, which is analogous to the one-broken-pair 
approximation employed for like-particle pairing, is analysed for various isovector and isovector-isoscalar
pairing Hamiltonians which can be solved exactly by diagonalisation.  

\end{abstract}

\maketitle

\section{Introduction}

The role played by  $\alpha$-like quartets for systems of nucleons  interacting by  proton-neutron (pn)
pairing forces has been debated for many years \cite{soloviev,bremond,flowers,eichler,dobes,chasman,zelevinsky}.
In a series of recent studies we have shown that $\alpha$-like quartets, defined as correlated structures of two protons and 
two neutrons coupled to total isospin $T=0$, represent the key elements for a proper description of  $N=Z$ systems governed by 
proton-neutron pairing interactions. In the case of a state-independent isovector pairing Hamiltonian, in particular, 
we have provided semi-analytical expressions of the $T=0$ seniority-zero eigenstates  and shown that these are linear superpositions 
of products of distinct $\alpha$-like quartets built by two collective $T=1$ pairs \cite{sasa_exact_t1}.
For the same Hamiltonian it has also been shown that a trial state formed by a single 
product of quartets of different structure provides ground state correlation energies which coincide with the 
exact values up to the 5th digit \cite{sasa_qm_qcm_t1}. Similar approximate solutions in terms of products
of distinct quartets have been also proposed for the isoscalar-isovector pairing interactions \cite{qm_t0t1}.

A particular class of quartet states of physical interest are those built by a product of identical quartets. 
These states, called quartet condensates, have been studied for both the isoscalar \cite{qcm_t1} and 
isoscalar-isovector \cite{qcm_t0t1,sasa_qcm_t0t1} pairing Hamiltonians in the framework of quartet 
condensation model (QCM) approach. In the special case of the state-independent isovector pairing Hamiltonian of Ref. \cite{sasa_exact_t1}, the link between the complex exact structure of the ground state  and this simple approximation scheme has been discussed in detail \cite{sasa_cond_t1}.
The QCM approach, which conserves exactly both the particle 
number and the isospin, has been found to describe accurately the ground state correlation energies of proton-neutron
pairing Hamiltonians, with an accuracy below $1\%$. The quartet correlations have turned out to be 
important also in the ground state of $N>Z$ systems. 
For these systems the ground state has been well approximated by a condensate of $\alpha$-like quartets to which 
a condensate of pairs, built with
the extra neutrons, is appended \cite{qcm_t1_ngz,qcm_t0t1_ngz}. Finally it is worth mentioning that also in the case of realistic shell-model type interactions, the quartet condensate has been found to approximate well the ground state of $N=Z$ nuclei \cite{hasegawa1,hasegawa2,qcm_sm_ex} and, to a good extent, also the first excited
$0^+$ states of $sd$-shell nuclei \cite{qcm_sm_ex}.

With the only exception of Ref. \cite{qcm_sm_ex}, all the
studies mentioned above have been fully addressed to a description of the ground states of proton-neutron pairing Hamiltonians.  The purpose of this paper is to extend these studies 
to the  excited states of these Hamiltonians. This work will be focused
on even-even $N=Z$ systems for which, as said above, the ground state can be  well-approximated 
by a condensate of $\alpha$-like quartets. For these systems we shall analyse a particular class of excited states
built by breaking a quartet from the condensate which describes the ground state and replacing it with an  ``excited " quartet. This approximation will be analyzed for various isovector and isovector-isoscalar
proton-neutron pairing Hamiltonians and the results will be contrasted with the exact eigenstates
provided by diagonalisation. 

The manuscript is structured as follows. In Section II, we will illustrate our
approach in the case of the isovector pairing. In Section III, we will discuss the case of an isovector plus isoscalar pairing Hamiltonian. Finally, in Section IV, we will summarize the results and draw the conclusions.

\section{Excited states for the isovector pairing}

The isovector pairing Hamiltonian considered in this section has the expression 

\begin{equation}
H=\sum_i  \epsilon_i N_i + 
\sum_{i,j} V^{T=1}_{J=0} (i,j) \sum_{T_z}P^+_{i,T_z} P^+_{j,T_z}
\end{equation}
where
\begin{equation}
{N}_i=\sum_{\sigma =\pm ,\tau =\pm\frac{1}{2}}a^\dag_{i\sigma\tau}a_{i\sigma \tau},~~~~
P^+_{i,T_z}= \sqrt{\frac{2j_i+1}{2}}[a^+_i a^+_i ]^{T=1,J=0}_{T_z}.
\end{equation}
The operator $a^\dag_{i\sigma\tau}$ ($a_{i\sigma\tau}$) creates (annihilates) a nucleon in the single-particle state $i$ characterized by the quantum numbers $(\sigma ,\tau )$, where $\sigma =\pm$ labels states which are conjugate with respect to time reversal and $\tau =\pm\frac{1}{2}$ is the projection of the isospin of the nucleon.
The operator $P^\dag_{iT_z}$ $(P_{iT_z})$ creates (annihilates) a pair of nucleons in time-reversed states with total isospin $T=1$. The three isospin projection $T_z$ correspond to $pp$, $nn$ and $pn$ pairs.
In Eq.(2) the pair operators are written for the case of a spherically-symmetric Hamiltonian with
pairs which have a well-defined angular momentum J=0. 
 
We start by recalling the quartet condensation model (QCM) for the ground state of this Hamiltonian,
which will be used below for introducing the new class of excited states. In Ref. \cite{sasa_qm_qcm_t1} it was
shown that the ground state of the Hamiltonian (1) with $n_q$/2 active protons and neutrons can be well approximated by a quartet condensate:
\beq
|QCM \rangle = (Q^+_{iv})^{n_q} | - \rangle
\eeq
where
\beq
Q^+_{iv}= \sum_{ij} x_{ij} [P^{\dag}_{i} P^{\dag}_{j}]^{T=0} = 
      \sum_{ij} x_{ij} \frac{1}{\sqrt{3}} (P^{\dag}_{i1}P^{\dag}_{j-1} + P^{\dag}_{i-1}P^{\dag}_{j1} - P^{\dag}_{i0}P^{\dag}_{j0})
\eeq
is the collective quartet built by a linear combination of two non-collective isovector pairs coupled to the
total isospin $T=0$. By construction the quartet (4) contains two types of 4-body correlations between the 
protons and neutrons: (a), those generated by the isospin coupling  and, (b), those arising from the the mixing parameters $x_{ij}$. 

In order to establish a connection between collective quartets and collective pairs, 
in Ref.\cite{qcm_t1} the mixing parameters have been taken separable in the indices,  i.e., $x_{ij}=x_i x_j$. 
In this approximation the ground state becomes
\beq
|\overline{QCM}\rangle = (\overline{Q}_{iv}^+)^{n_q} | - \rangle
\eeq
where the new quartet operator
\beq
\overline{Q}_{iv}^+= 2\Gamma^+_1\Gamma^+_{-1} -(\Gamma^+_0)^2
\eeq
is expressed in terms of the collective pair $\Gamma^+_t=\sum x_i P^+_{it}$. From Eq. (6) one can see that
in this approximation the quartets contains only those 4-body correlations generated by the isospin coupling.
We remark that it has been recently shown that the QCM state (5) results from the projection on the isospin $T=0$ and the particle number of the BCS-type function  $e^{\Gamma^+_0}|- \rangle $ \cite{baran}.

In order to study the excitation spectrum of the Hamiltonian (1) for the same system of protons and neutrons,
in the present study we shall consider a new class of QCM states obtained by removing a quartet from the condensate
describing the ground state and replacing it with a new ``excited" quartet. We shall explore this approximation in correspondence with both types of condensates (3) and (5).

We shall begin from the condensate (3) and refer to this case as Approximation (A).
The excited  states have the form 
\beq
|\Phi_\nu \rangle = \tilde{Q}_\nu^+ (Q^+_{iv})^{n_q-1} |-\rangle ,
\eeq
where 
\beq
\tilde{Q}_{\nu}^+ = \sum_{ij} y^{(\nu)}_{ij} [P^+_i P^+_j]^{T=0}
\eeq
represents the excited collective quartet. These excited states are therefore linear superpositions of the states
\beq
[P^+_i P^+_j]^{T=0} (Q^+_{iv})^{n_q-1} |-\rangle .
\eeq
In order to construct the amplitudes $y^{(\nu)}_{ij}$ defining the collective quartet $\tilde{Q}_{\nu}^+$, once a QCM calculation for the ground state has been performed  and the quartet $Q^+_{iv}$ has been defined, it suffices to diagonalize
the Hamiltonian (1) in the space spanned by the non-orthogonal states (9).
Being built in terms of  non-collective operators $P^+_{iT_z}$ which create pairs
of nucleons in time-reversed states, the eigenstates (7) are zero seniority states \cite{richardson}. 


\begin{figure*}[ht]
\begin{center}
\hbox{\includegraphics[width=0.40\textwidth, angle=-90]{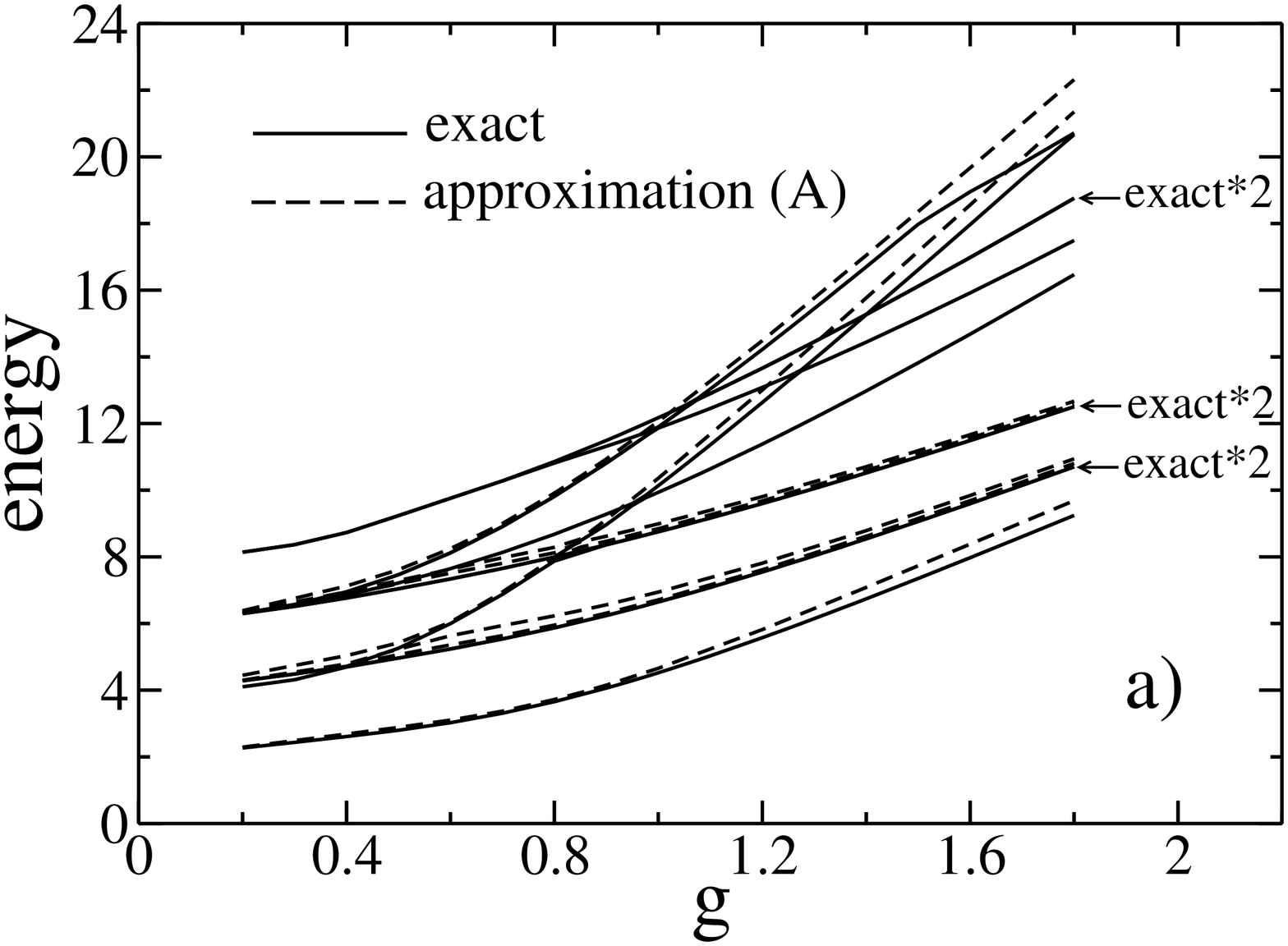}\quad
\includegraphics[width=0.40\textwidth, angle=-90]{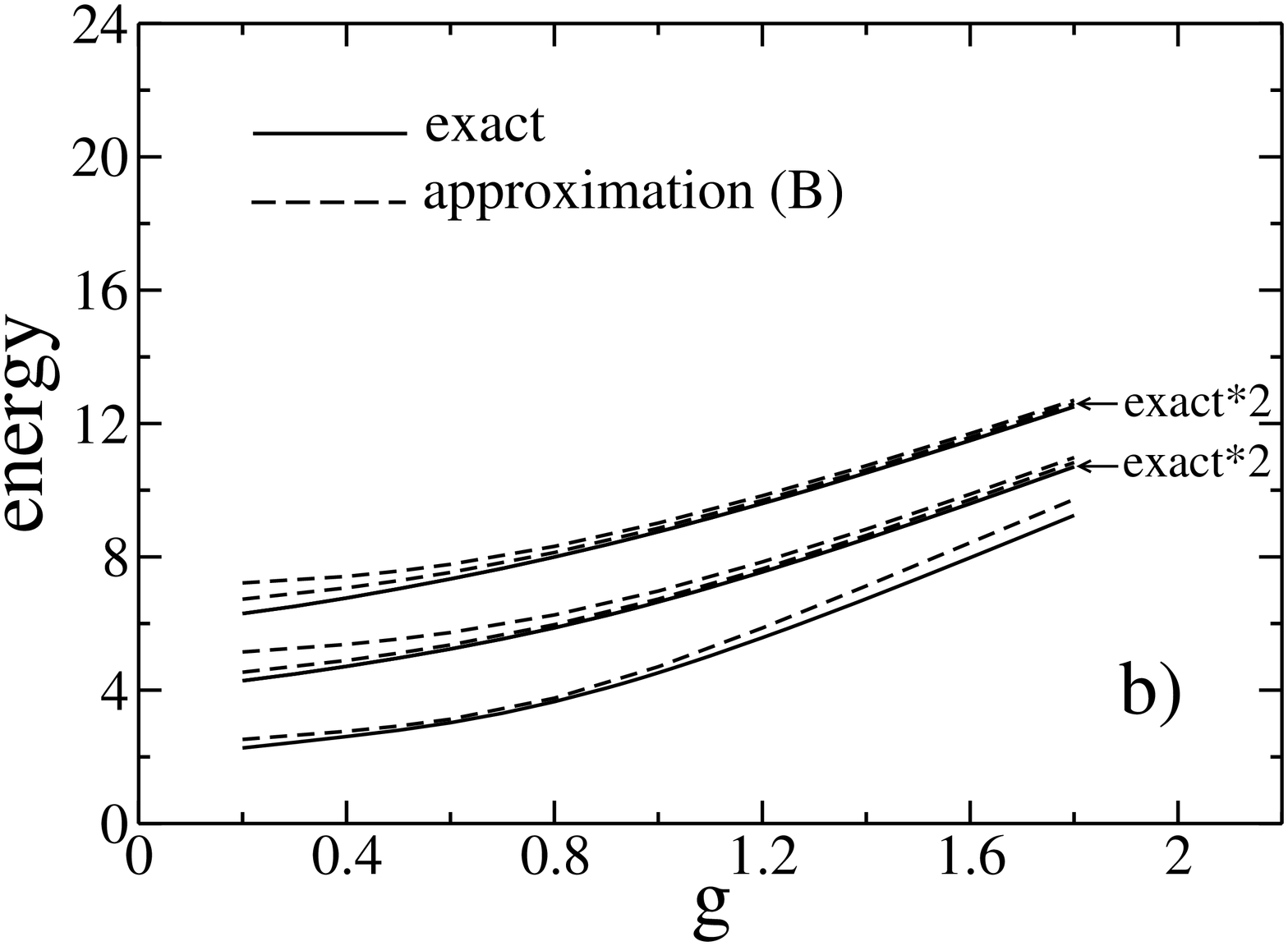}}
\end{center}
\caption{Excitation spectra of the isovector pairing Hamiltonian (1) for a system of $N=Z=6$ particles
moving on 6 equidistant levels. Dashed lines in Figs. 1a and 1b refer, respectively, to the approximations (A) and (B) discussed in the text while full lines represent the exact results. Energies and the pairing strength $g$ are in units of the spacing $\Delta\epsilon$ between the levels.}
\end{figure*}

To test this approximation, we shall consider a system with 6 protons and 6 neutrons interacting trough a state independent isovector pairing force (i.e. $V^{T=1}_{J=0} (i,j)\equiv -g$ in Eq. (1)) and distributed over 6 equidistant levels with four-fold degeneracy (due to the presence of both spin and isospin degrees of freedom).
There are two different ways (equivalent in practice) to interpret this model space. On one side , this space 
can be associated with
a set of single-particle states of orbital angular momentum $l$=0 and $j$=1/2. In this case the quartets
are built by pairs with angular momentum $J$=0 and, consequently, all the states (7) have $J$=0.
Alternatively, the single-particle levels can represent a set of axially-deformed single-particle states
associated with an intrinsic deformed mean field. In the latter case the pairs operators, defined by 
$P^+_{i,T_z}= [a^+_i a^+_i ]^{T=1}_{T_z}$, and the eigenstates (7) have $J_z$=0 but not a well-defined angular momentum. Following Ref. \cite{sasa_exact_t1}, we have adopted the single particle energies $\epsilon_i=-16+2(i-1)$ which are characterized by a constant spacing $\Delta\epsilon =2$.

In Fig. 1a we compare the excitation energies provided by the approximation (7), as a function of the
pairing strength $g$, with the exact results obtained by diagonalisation. 
One  can observe that the approximation (7) works well for all pairing strengths, from weak to strong coupling regimes. It can be also noticed that the exact low-lying spectrum contains a few states which cannot be represented by the approximation (7).
      
As a next step we shall consider the same type of approximation discussed so far but in correspondence with the 
ground state condensate (5) where all quartets are built in terms of the isovector collective pair ${\Gamma}^+$. We shall refer to this case as Approximation (B). The excited states are now defined as 
\beq
|\overline{\Phi}_\nu \rangle = \hat{Q}_\nu^+ (\overline{Q}^+_{iv})^{n_q-1} |-\rangle ,
\eeq
with
\beq
\hat{Q}_{\nu}^+ = [\tilde{\Gamma}^+_{\nu} \Gamma^+]^{T=0} 
\propto \tilde{\Gamma}^+_{\nu,1} \Gamma^+_{-1} + \tilde{\Gamma}^+_{\nu,-1} \Gamma^+_{1} 
- \tilde{\Gamma}^+_{\nu,0} \Gamma^+_{0}.
\eeq
The state $|\overline{\Phi}_\nu \rangle$ differs from the corresponding ground state $|\overline{QCM}\rangle$ only for the presence of the ``excited" pair $\tilde{\Gamma}^+_{\nu,t}=\sum_i z^{(\nu)}_i P^+_{it}$,
the pair ${\Gamma}^+$ being instead that defining the quartets $\overline{Q}^+_{iv}$. In order to define the coefficients $z^{(\nu)}_i$ it suffices to diagonalize the Hamiltonian in the basis of non-orthogonal states 
\beq
[P^+_i \Gamma^+]^{T=0} (\overline{Q}^+_{iv})^{n_q-1} |-\rangle.
\eeq
 
In Fig. 1b we show the eigenvalues corresponding to the excited quartets (10) for the same system considered above.
Only 5 approximate excited states can be built in this case (the index $i$ of (12) ranging over the number of the levels) and they are seen to follow quite closely the behavior of 5 exact low-lying excited states. From a comparison with Fig. 1a one may notice that, for values of $g\agt 0.8$, the 5 exact eigenstates in this figure coincide with the 5 lowest excited states of the Hamiltonian (1) while for smaller values of $g$ an ``intruder" exact eigenstate exists which crosses these states and which is not reproduced in the Approximation (B). In this figure, for simplicity, only 5 exact excited eigenstates have been reported and the agreement with the approximate ones appears fairly good, the largest deviations being observed in the weak coupling regime.  

In order to better understand the quality of the Approximations (A) and (B) in the calculations just discussed, in Fig. 2 we show a more detailed description of the results of these approximations in a specific case. The calculations of this figure  refer to a value of the strength $g=1.0$ and report not only the spectra but also the overlaps between exact and approximate eigenstates. One can notice that the overlaps are very large both in the approximation (A) and (B). An overlap equal to zero indicates that the corresponding
exact eigenstate is not a QCM state. Two remarks are in order with reference to this figure. The first remark concerns the approximate ground states. These ground states are those resulting from the diagonalization of the Hamiltonian (1) in the space of states (9) (Approximation (A)) and in the space of states (12) (Approximation (B)). Strictly speaking, thus, they are not true QCM condensates since one of the quartets results from a diagonalization and is not constrained to be equal to the others. However, the fact that the QCM ground state corresponds to a minimum in energy, causes this new quartet to be essentially identical to the others as we have also verified by the fact that both the energy and the overlap of this state are basically indistinguishable from those of the true QCM state. The second remark  has to do with a peculiarity of the exact spectrum already evidenced in Figs. 1a and 1b, namely the existence of degeneracies. The evaluation of the overlaps between a generic state $|\alpha\rangle$ and two denenerate states $|\Psi_1\rangle$ and $\Psi_2\rangle$ is hampered by the fact that the wave functions of the degenerate states cannot  be unambiguously defined since any other two states $|\Psi^{(+)}_{12}\rangle =d_1|\psi_1\rangle + d_2|\psi_2\rangle$ and
$|\Psi^{(-)}_{12}\rangle =d_1|\psi_1\rangle - d_2|\psi_2\rangle$, with $d^2_1+d^2_2=1$, also represent a pair of degenerate eigenstates with the same energy. The overlaps $\langle\alpha |\Psi^{(\pm )}_{12}\rangle$ obviously depend on the (arbitrary) coefficients $d_1$ and $d_2$. In such a circumstance we have followed the approach of Ref. \cite{michelangelo_overlap} and introduced the quantity $M_\alpha^{(12)}=\langle\alpha |\Psi_1\rangle^2 +\langle\alpha |\Psi_2\rangle^2$. This quantity is invariant with respect to any transformation $|\Psi^{(\pm )}_{12}\rangle$ and it can be seen to provide the maximum squared overlap between the state $|\alpha\rangle$ and a generic state $|\Psi^{(+)}_{12}\rangle$. This maximum is found in correspondence with the state 
\beq
|\Psi^{(+)}_{12}\rangle =\frac{1}{\sqrt{M_\alpha^{(12)}}}(\langle\alpha |\Psi_1\rangle |\Psi_1\rangle +
\langle\alpha |\Psi_2\rangle |\Psi_2\rangle )
\eeq
while the paired eigenstate 
\beq
|\Psi^{(-)}_{12}\rangle =\frac{1}{\sqrt{M_\alpha^{(12)}}}(\langle\alpha |\Psi_1\rangle |\Psi_1\rangle -
\langle\alpha |\Psi_2\rangle |\Psi_2\rangle )
\eeq
is, by construction, such that $\langle\alpha |\Psi^{(-)}_{12}\rangle =0$ \cite{michelangelo_overlap}. The overlap shown in Fig. 2 in correspondence to two degenerate states $|\Psi_1\rangle$ and $\Psi_2\rangle$ is thus the square root of the quantity $M_\alpha^{(12)}$.

\begin{figure}
\includegraphics{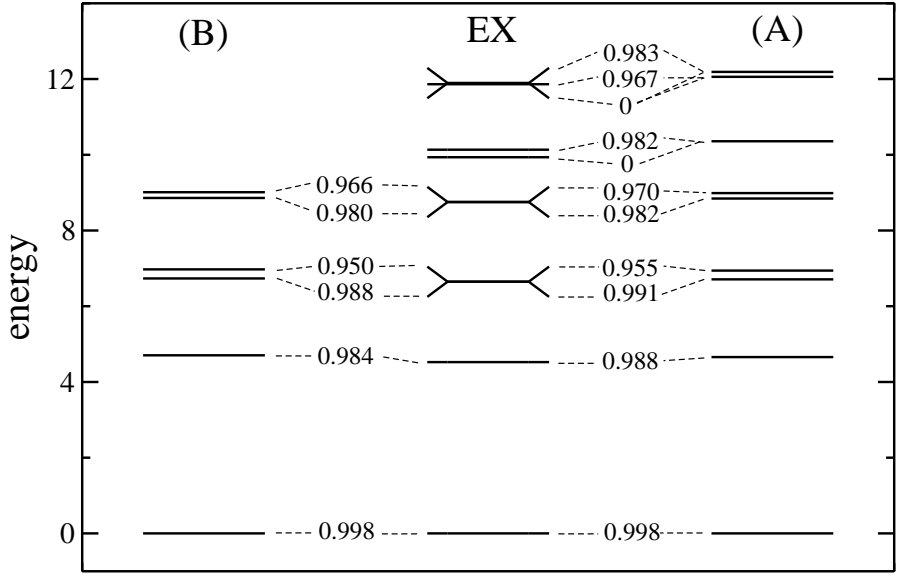}
\mbox{}\\[9.0cm]
\caption{Low-lying spectra of the isovector pairing Hamiltonian (1) for the same system discussed in Fig. 1
and a pairing strength $g=1$. The spectra (A) and (B) refer, respectively, to the Approximations (A) and (B)
discussed in the text while the spectrum EX corresponds to the exact one. Energies and the pairing strength $g$ are in units of the spacing $\Delta\epsilon$ between the levels.}
\end{figure}

The examples discussed so far have involved quartets formed by the isovector operators $P^+_{iT_z}$, which under the assumption of spherical symmetry, are characterized by an angular momentum $J=0$. In what follows, aiming at a more realistic application of the isovector pairing Hamiltonian (1) in a spherical mean field, we introduce the most general pair creation operator 
\beq
P^+_{JJ_z,TT_z}(i,j)=[a^+_ia^+_j]^{JT}_{J_zT_Z}
\eeq
and, by means of this, the most general collective $T=0$ quartet
\beq
\tilde{Q}^+_{\nu ,JJ_z}=\sum_{T'}\sum_{J_1(i_1j_1)}\sum_{J_2(i_2j_2)}Y^{(\nu )}_{JJ_z}(T',J_1(i_1j_1),J_2(i_2j_2))
[P^+_{J_1,T'}(i_1,j_1)P^+_{J_2,T'}(i_2,j_2)]^{J,T=0}_{J_z}.
\eeq
This quartet is employed to define the exited states 
\beq
|\Phi_{\nu, JJ_z}\rangle = \tilde{Q}_{\nu ,JJ_z} (Q^+_{iv})^{n_q-1} |-\rangle ,
\eeq
where the collective quartet $Q^+_{iv}$ is still restricted to isovector pairs only.
$Q^+_{iv}$ has been assumed to be of the type (4), by therefore
excluding a factorization of the coefficients $x_{ij}$.
Similarly to the cases discussed above,
in order to find the coefficients $Y^{(\nu )}_{JM}$ of the quartet (16) and so construct the excited states
one needs to diagonalize the Hamiltonian (1) in the space of non-orthogonal states
\beq
[P^+_{J_1,T'}(i_1,j_1)P^+_{J_2,T'}(i_2,j_2)]^{J,T=0}_{J_z}(Q^+_{iv})^{n_q-1} |-\rangle .
\eeq

\begin{figure}
\includegraphics{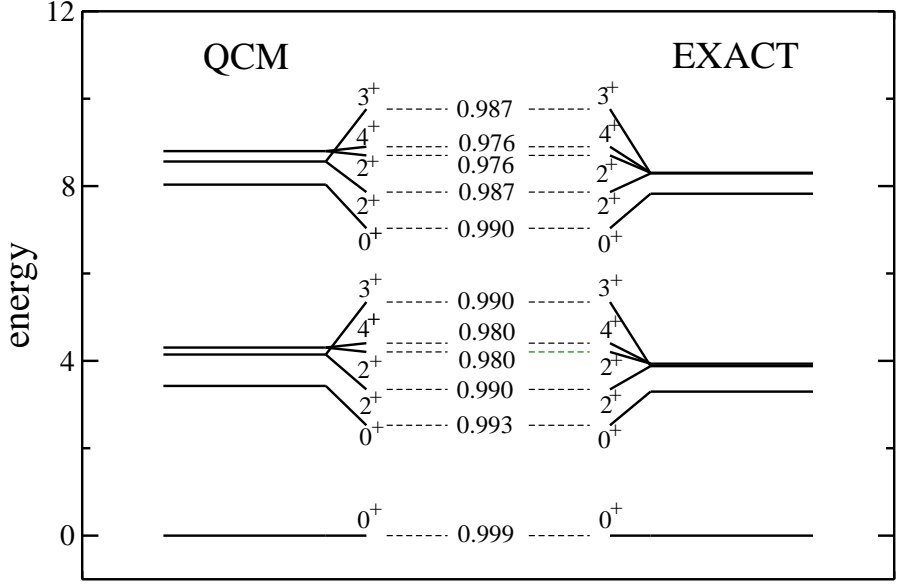}
\mbox{}\\[9.0cm]
\caption{The low-lying spectrum provided by the QCM approximation (17) for the valence nucleons of $^{28}$Si 
interacting by an isovector pairing force extracted from the USDB interaction. The numbers are the
overlaps between the QCM and the exact wave functions. Energies are in MeV.}
\label{fig3}
\end{figure}

To asses the validity of the approximation (17) we have performed calculations for a system of $N=Z=6$
nucleons moving in the $sd$-shell and interacting with an isovector pairing force extracted from
the USDB interaction \cite{usdb}. This system corresponds to $^{28}$Si. 
The energies obtained for the low-lying $T=0$ states (17) are given in Fig.3 and are compared with the
exact eigenvalues ( calculated with the shell model code BIGSTICK \cite{bigstick}). As it can be seen, 
the approximation (17) reproduces quite well the exact results with overlaps between corresponding states
which are close to unity for all the low-lying states. 

\section{Excited states for the isovector-isoscalar pairing}

The isovector-isoscalar pairing Hamiltonian has the expression 
\begin{equation}
H=\sum_i  \epsilon_i N_i + 
\sum_{i,j} V^{T=1}_{J=0} (i,j) \sum_{T_z}P^+_{i,T_z} P_{j,T_z}+
\sum_{i\leq j,k\leq l} V^{T=0}_{J=1}(ij,kl) 
\sum_{J_z}D^+_{ij,J_z} D_{kl,J_z}.
\end{equation}
The first two terms are the same as in Eq. (1) while the last term is the isoscalar pairing interaction
written in term of the isoscalar pair operator
\begin{equation}
D^+_{j_1 j_2 J_z}= \frac{1}{\sqrt{1+\delta_{j_1j_2}}}[a^+_{j_1} a^+_{j_2} ]^{J=1,T=0}_{J_z}
\end{equation} 

As in the previous section, we start by recalling the QCM approach for the ground state of the
isovector-isoscalar Hamiltonian \cite{sasa_qcm_t0t1}. For even-even $N=Z$ systems the QCM ansatz
for the ground state has formally the same expression as in the case of isovector pairing 
\begin{equation}
|\Psi_{gs}\rangle  = (Q^+_{ivs})^{n_q} |0 \rangle .
\end{equation}
The difference is that now the quartet operator $Q^+_{ivs}$, still having total isospin $T=0$, is the sum of two  quartets
\beq 
Q^+_{ivs}=Q^+_{iv}+Q^+_{is},
\eeq
where $Q^+_{iv}$ is the quartet (4) built by isovector pairs while $Q^+_{is}$ is
formed by  two isoscalar pairs coupled to total $J=0$, i.e.,
\beq
Q^+_{is}= \sum_{j_1 j_2 j_3 j_4} y_{j_1 j_2 j_3 j_4} [D^+_{j_1 j_2}D^+_{j_3 j_4}]^{J=0}.
\eeq
A simpler version of this approach can be obtained by adopting in (23) the factorisation
$y_{j_1 j_2 j_3 j_4} = y_{j_1,j_2=\bar{j}_1} y_{j_3,j_4=\bar{j}_3}$ (the bar indicating time-reversing) and by using the expression 
(6) for the isovector quartet. This QCM approximation has been investigated in detail in Ref. \cite{qcm_t0t1} and will not be further
discussed in the present work.

Acting as in the isovector pairing case, 
in correspondence with the QCM ansatz (21) for the ground state, we  construct a class of excited states by
replacing a quartet of the condensate with an ``excited" quartet. For
the  case of a spherically-symmetric mean field, these states take the form 
\beq
|\Phi_{\nu, JJ_z}\rangle = \tilde{Q}_{\nu ,JJ_z} (Q^+_{ivs})^{n_q-1} |-\rangle ,
\eeq
where the operator $\tilde{Q}_{\nu ,JJ_z}$ is identical to that defined in Eq. (16). In order to define its coefficients
$Y^{(\nu )}_{JJ_z}$, one has now to diagonalize the Hamiltonian (19) in the basis
of non-orthogonal states
\beq
[P^+_{J_1,T'}(i_1,j_1)P^+_{J_2,T'}(i_2,j_2)]^{J,T=0}_{J_z}(Q^+_{ivs})^{n_q-1} |-\rangle .
\eeq

\begin{figure}
\includegraphics{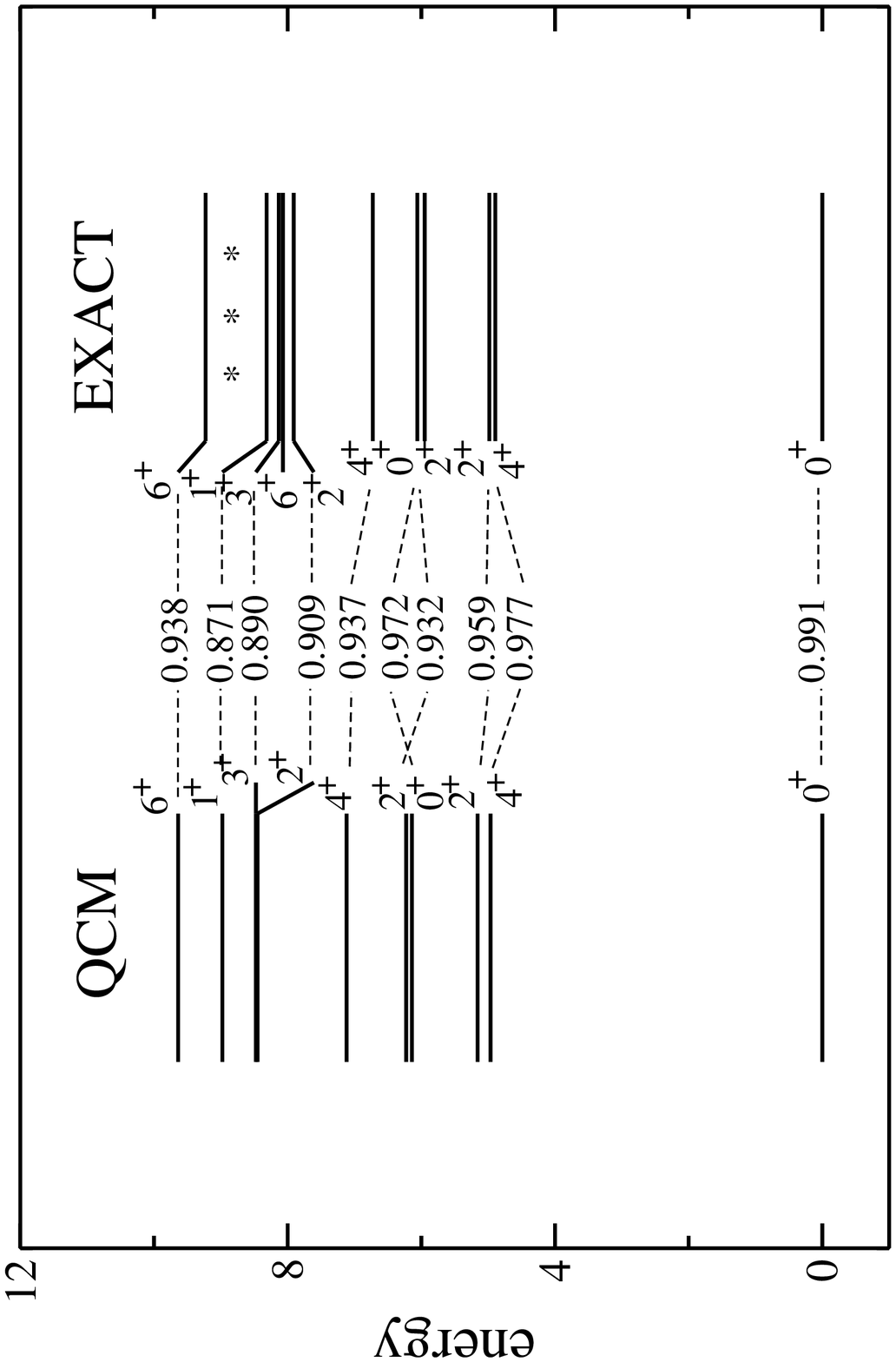}
\mbox{}\\[9.0cm]
\caption{The low-lying spectrum provided by the QCM approximation (24) for the valence nucleons of $^{28}$Si 
interacting by an isovector-isoscalar pairing force extracted from the USDB interaction. The numbers are the
overlaps between the QCM and the exact wave functions. Energies are in MeV.}
\label{fig4}
\end{figure}

To illustrate the accuracy of the approximation (24) we have still referred to the case of
$^{28}$Si and assumed an isovector-isoscalar pairing force corresponding to
the $(J=0,T=1)$ and $(J=1,T=0)$ channels of the the USDB interaction \cite{usdb}. Exact and approximate spectra are shown in Fig. 4. It can be seen that the inclusion of the isoscalar force removes the
degeneracies observed in the case of the isovector interaction. The overall agreement is good also in this case
although the quality of the overlaps is, in some cases, not as high as that of Fig. 3.
As a peculiarity, we notice that the first excited $J=6$ state has not a corresponding state in the QCM approximation while the second $J=6$ exact state is well reproduced both for the energy and  the overlap.

\section{Summary and conclusions}

We have extended the quartet condensation model (QCM) to describe the $T=0$ excited states of proton-neutron pairing Hamiltonians . These excited states have been generated by breaking a quartet from the quartet condensate which describes the ground state and replacing it with an ``excited" quartet. We have first discussed this approach for a state-independent isovector pairing force acting on a system of $N=Z=6$ nucleons
moving on a set of equidistant level. For such a system we have considered two levels of approximation,
one assuming the quartets of the ground state formed by two identical isovector collective pairs coupled to $T=0$  and the other, less restrictive, letting the quartets be simply superpositions of products of two uncorrelated isovector pairs coupled to total isospin $T=0$. In the first case, the ``excited" quartet has been generated by breaking only one of the two collective pairs. In both cases a very good agreement between exact and approximate spectra has been found both at the level of the energies and of the overlaps.

As further applications we have considered the cases of isovector and isovector-isoscalar pairing Hamiltonians
in a spherical mean field. The quartets of the ground state condensates have been restricted in these two cases to isovector and isovector plus isoscalar pairs only while the
excited quartet has been assumed in both cases to be the most general combination of two protons and two neutrons coupled to total isospin $T$=0 and total angular momentum $J$. We have examined, in
particular, the case of $N=Z=6$ nucleons moving in the $sd$-shell (corresponding to $^{28}$Si) with isovector and
isovector-isoscalar pairing forces extracted from the shell-model interaction USDB \cite{usdb}.
In both cases the low-lying excited states predicted by the extended QCM approach have compared well with
the exact eigenstates. As a major result, then, all the results illustrated in this paper clearly point to the relevance 
that $\alpha$-like degrees of freedom play not only in the ground state but also in the excited states of proton-neutron pairing Hamiltonians. 
The approach illustrated in this paper provides an effective tool to construct approximate spectra of these Hamiltonians by allowing a simple 
intepretation of the structure of their eigenstates.

We like to conclude by noticing the interesting analogy between the eigenstates of 
proton-neutron and like-particle paring Hamiltonians. As already pointed out in previous
studies, the QCM ansatz for the ground state of even-even N=Z systems is the analogous of the 
particle-number projected-BCS (PBCS) approximation for like-particle systems proposed many years
ago by Bayman and Blatt \cite{bayman,blatt}. On the other hand, the one-broken-quartet approximation 
for the excited states of $N=Z$ systems discussed in the present work shows a clear analogy with the one-broken-pair approximation employed for the treatment the excited states in like-particle systems \cite{talmi}.

\vskip 0.3cm
\begin{acknowledgments}
This work was supported by a grant of Romanian Ministry of Research and Innovation, CNCS - UEFISCDI, project number PCE 160/2021, within PNCDI III.
\end{acknowledgments}

\end{document}